\documentclass{aa}
\usepackage{graphics,epsf,epsfig,rotating,times}
\begin{document}

   \thesaurus{11     
              (11.01.2;  
               11.14.1;  
               11.02.2;  
               13.07.2)  
               }
\title{Very High Energy Gamma-ray spectral properties of Mkn~501 from
  CAT \v Cerenkov telescope observations in 1997
  }

\author{
A. Djannati-Ata\"{\i} \inst{7} \and
F. Piron \inst{4} \and 
A. Barrau \inst{5}\fnmsep\thanks{\emph{Present address:} Institut des Sciences 
Nucl\'eaires, 53 avenue des Martyrs, F-38026 Grenoble cedex, France} \and 
L. Iacoucci \inst{4} \and 
M. Punch \inst{7} \and
J.-P. Tavernet \inst{5} \and
R. Bazer-Bachi \inst{2} \and 
H. Cabot \inst{3} \and          
L.-M. Chounet \inst{4} \and 
G. Debiais \inst{3} \and 
B. Degrange \inst{4} \and 
J.-P. Dezalay \inst{2} \and 
D. Dumora \inst{1} \and 
P. Espigat \inst{7} \and
B. Fabre \inst{3} \and 
P. Fleury \inst{4} \and 
G. Fontaine \inst{4} \and 
C. Ghesqui\`ere \inst{7} \and
P. Goret \inst{8} \and
C. Gouiffes \inst{8} \and
I.A. Grenier \inst{8}\fnmsep\inst{9} \and
S. Le Bohec \inst{4}\fnmsep\thanks{\emph{Present address:} 
Physics \& Astronomy Department, Iowa State University, Ames, IA 50011, USA}\and 
I. Malet \inst{2} \and 
C. Meynadier \inst{3} \and 
G. Mohanty \inst{4} \and 
E. Nuss \inst{3} \and 
E. Par\'e \inst{4} \and 
J. Qu\'ebert \inst{1} \and 
K. Ragan \inst{1}\fnmsep\thanks{\emph{Present address:} 
Physics Department, McGill University, Montreal, H3A 2T8, Canada} \and
C. Renault \inst{5} \and
M. Rivoal \inst{5} \and
L. Rob \inst{6} \and
K. Schahmaneche \inst{5} \and
D.A. Smith \inst{1} 
}

\institute{
Centre d'Etudes Nucl\'eaire de
Bordeaux-Gradignan, France (IN2P3/CNRS) 
\and
Centre d'Etudes Spatiales des
Rayonnements, Toulouse, France (INSU/CNRS)
\and
Groupe de Physique Fondamentale,
Universit\'e de Perpignan, France (IN2P3/CNRS)
\and
Laboratoire de Physique Nucl\'eaire des Hautes Energies,
Ecole Polytechnique, Palaiseau, France (IN2P3/CNRS)
\and
Laboratoire de Physique Nucl\'eaire et de Hautes Energies,
Universit\'es de Paris VI/VII, France (IN2P3/CNRS)
\and
Nuclear Center, Charles University, Prague, Czech Republic
\and
Physique Corpuscul\-aire et Cosmologie,
Coll\`ege de France, Paris, France (IN2P3/CNRS)
\and
Service d'Astrophysique, Centre d'Etudes
de Saclay, France (DAPNIA/CEA)
\and
Universit\'e Paris VII, France
}

\offprints{A. Djannati-Ata\"\i}
\mail{djannati@cdf.in2p3.fr}

\date{Received May 26, 1999; accepted August 16, 1999}

\titlerunning{VHE Spectral properties of Mkn 501 with CAT in 1997}
\maketitle

\begin{abstract}

The BL Lac object Mkn~501 went into a very high state of activity during 1997,
both in VHE $\gamma$-rays and X-rays.  We present here results from
observations at energies above $250\:\mathrm{GeV}$ carried out between March
and October 1997 with the C{\small AT} \v Cerenkov imaging Telescope.

The average differential spectrum  between $330\:\mathrm{GeV}$ and 
$13\:\mathrm{TeV}$ shows significant curvature and is well represented by 
$\phi_0 E_\mathrm{TeV}^{-(\alpha + \beta\log_{10}\!E_\mathrm{TeV})}$, with:

\indent $\phi_0 = \mathrm{ 5.19 \pm 0.13^{stat} \pm 0.12^{sys-MC} }\\
\indent \indent \indent \;\;\;\;
\stackrel{+1.66}{_{-1.04}}^\mathrm{sys-atm}
\times 10^{-11}\:\mathrm{cm^{-2}s^{-1}TeV^{-1}}$, \\
\indent$\alpha \:\, = 2.24 \pm \mathrm{ 0.04^{stat} \pm 0.05^{sys} }$, and\\
\indent$\beta \hspace{4.2pt} = 0.50 \pm 0.07^\mathrm{stat} $ (negligible systematics).

The TeV spectral energy distribution of Mkn~501 clearly peaks in the range
$500\:\mathrm{GeV}$--$1\:\mathrm{TeV}$.
Investigation of spectral variations shows a significant hardness-intensity
correlation with no measurable effect on the curvature. This can be
described as an increase of the peak TeV emission energy with intensity.

Simultaneous and quasi-simultaneous C{\small AT} VHE $\gamma$-ray and
BeppoS{\small AX} hard X-ray detections for the highest recorded flare on
$16^\mathrm{th}$  April and for lower-activity states of the same period
show correlated variability with a higher 
luminosity in X-rays than in $\gamma$-rays.

The observed spectral energy distribution and the correlated
variability between X-rays and $\gamma$-rays,
both in amplitude and in hardening of spectra, favour a two-component
emission scheme where the 
low and high energy components are attributed to synchrotron and inverse
Compton (IC) 
radiation, respectively.

\end{abstract}

\begin{figure*}
\leavevmode
\begin{center}
\hbox{
\hspace{0pt}
\epsfig{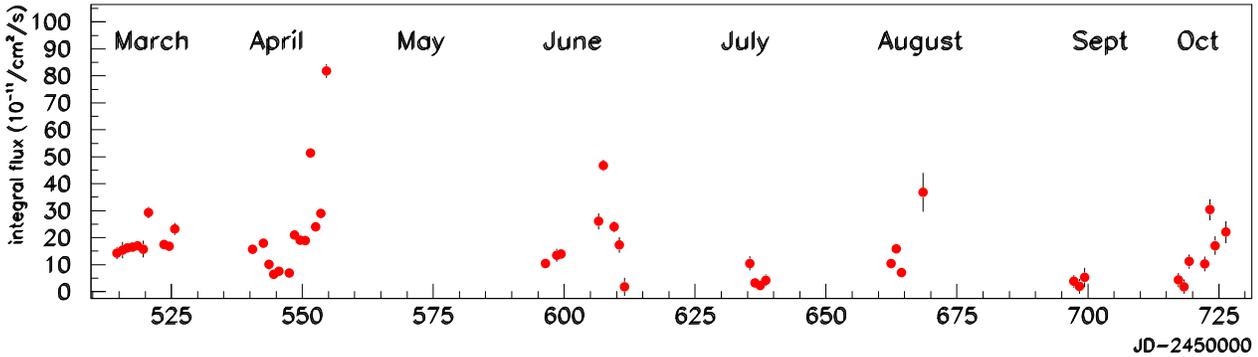}
}
\caption{
Mkn~501 nightly integral flux levels above  $250\:\mathrm{GeV}$,
expressed in units of 
$10^{-11}\:\mathrm{cm^{-2}s^{-1}}$, for observations in 1997. The
average spectral shape as derived in Sect.~4 has been assumed 
in order to estimate the integral flux for observations far from the Zenith.
}
\label{figlctot} 
\end{center}
\end{figure*}
\noindent  This article is dedicated to the memory of Eric Par\'e, one of the 
  originators of C{\small AT}, who died in a road accident in July, 1998.

\section{Introduction}

Mkn~501, at $z=0.034$, is the second-closest BL Lac object and, with 
Mkn~421 ($z=0.031$), is one of the two extra-galactic sources thus far 
detected and confirmed at TeV energies.
Its TeV emission was discovered in 1995 by the Whipple observatory group
(\cite{quinn96}) above $300\:\mathrm{GeV}$, at the level of 8\% of 
the TeV flux of the Crab Nebula, and subsequently confirmed by the
H{\small EGRA} collaboration (\cite{bradbury97}). 

Initially, Mkn~501 had been placed in the radio-selected BL Lac family 
({\cite{stickel91}, from 1~Jy BL Lac and S4 radio samples). 
However, its spectral energy distribution (SED) 
in recent years resembled typical High-Frequency
BL Lac objects (mostly X-ray selected) with the low-energy peak 
in the EUV/soft X-ray energy band, making it a bright Einstein Slew
Survey source (\cite{ciliegi95}). In fact, it can be considered  as being 
intermediate between the radio- and X-ray bright classes of BL Lac 
(\cite{lamer98}).

From March to October 1997, Mkn~501 showed
dramatic activity in the VHE $\gamma$-ray band, exhibiting flares
during which it became up to $\sim\!\!\!8$ times brighter than the Crab
Nebula (e.g., see \cite{protheroe97}); i.e., two orders of magnitude
brighter than its discovery level. 
 
During the same period, Mkn~501 was also extremely active in the X-ray band
as reported by the All Sky Monitor (ASM) on board the {\it Rossi X-Ray Timing
Explorer} (RXTE). 
In April, high-intensity flares were recorded in the hard
X-ray range by RXTE (\cite{lamer98}), BeppoS{\small AX}  (\cite{pian98}),
and O{\small SSE} (\cite{catanese97}) on board the {\it Compton Gamma
Ray Observatory} (CGRO). The synchrotron peak underwent a shift towards high
energies as compared to archival data, shifting to energies of
$100\:\mathrm{keV}$.
This unprecedentedly high energy made Mkn~501 the most extreme
known BL Lac object in the X-ray sky.

This paper is dedicated to the study of detailed spectral properties of 
this source, based on observations carried out with the C{\small AT}
\v Cerenkov imaging telescope
above $250\:\mathrm{GeV}$ in 1997. 

The next section describes briefly the C{\small AT} telescope and its specific
event reconstruction method. The data-sample used for this paper and the
resulting source light-curve are presented in Sect.~3. 
Spectra derived for different-activity states are given in Sect.~4 and 
intensity-hardness correlations are studied in Sect.~5.
In Sect.~6 we compare simultaneous and quasi-simultaneous C{\small AT} and 
BeppoS{\small AX} data for the highest flare recorded on April 
$16^\mathrm{th}$ to the lower activity states of the $7^\mathrm{th}$ and
$11^\mathrm{th}$ of the same period. 
In Sect.~7 we discuss results and their implications
for proposed emission models. 
The conclusions are given in Sect.~8.

\section{C{\small AT} \v Cerenkov Imaging Telescope}

The $17.8\:\mathrm{m}^2$  C{\small AT} (\v Cerenkov Array at Th{\'e}mis)
imaging telescope started operation in Autumn 1996 on the site of the former
solar plant Th{\'e}mis (France).
A complete description of the C{\small AT} telescope can be
found in \cite{catdet98}. 

The instrument records \v Cerenkov light from the particles in
cosmic-ray showers. The  
analysis of the shape and the pointing of the resulting images, as discussed in
\cite{lebohec98}, takes full advantage of the very-high-definition camera.
This latter has a $4.8\degr$ full field of view
which is comprised of a central region of 546 phototubes in a hexagonal matrix
spaced by $0.13\degr$ and 54 surrounding tubes in two ``guard rings''.

The $\gamma$-ray image analysis is based on the comparison of
individual events and theoretical average images as a function of
impact parameter and energy.
Accurate analysis of the longitudinal light profile 
and lateral width
of the shower
image is possible thanks to the high-resolution camera. This fitting
procedure allows the
direction and energy of each $\gamma$-ray to be determined with good
accuracy: the angular resolution ($0.14\degr$)
is of the order of the pixel size; when selecting showers with an impact 
parameter less than $130\:\mathrm{m}$, the energy resolution is about 20\%,
independent of energy, from approximately $300\:\mathrm{GeV}$ 
to $15\:\mathrm{TeV}$. 
At the highest energies, the image tails are cut by the field of view,
but the light-profile analysis still permits a good energy estimation.

It should be noted that, in general, for the Atmospheric \v Cerenkov Technique,
there is a systematic uncertainty due to variations  in the
transparency of the atmosphere. 
Additional uncertainties due to unmastered detector drifts exist, but are 
fairly low for C{\small AT}, and are monitored within the data (e.g., using
muon ring images). A conservative estimate of the overall uncertainty
on the absolute energy scale is $20\%$.

To separate $\gamma$-ray initiated shower images from those due to
hadronic showers we use the $\chi^2$-like goodness-of-fit parameter given by the
above procedure, requiring the probability $P_{\chi^2} > 0.35$, and the
pointing angle (which is the angle at 
the image barycentre between the actual source position and the
reconstructed angular origin of the $\gamma$ ray), $\alpha < 6\degr$.
In addition, only images with more than 30 photo-electrons and whose
fourth brightest pixel has more than 3 photo-electrons are selected, in 
order to eliminate the lowest-energy triggers.
These cuts yield a rejection factor of about 200, with a
$\gamma$-ray efficiency better than 40\%. Further details on this
specific method can be found in \cite{lebohec98}. 

\section{Data Sample}
\begin{figure*}
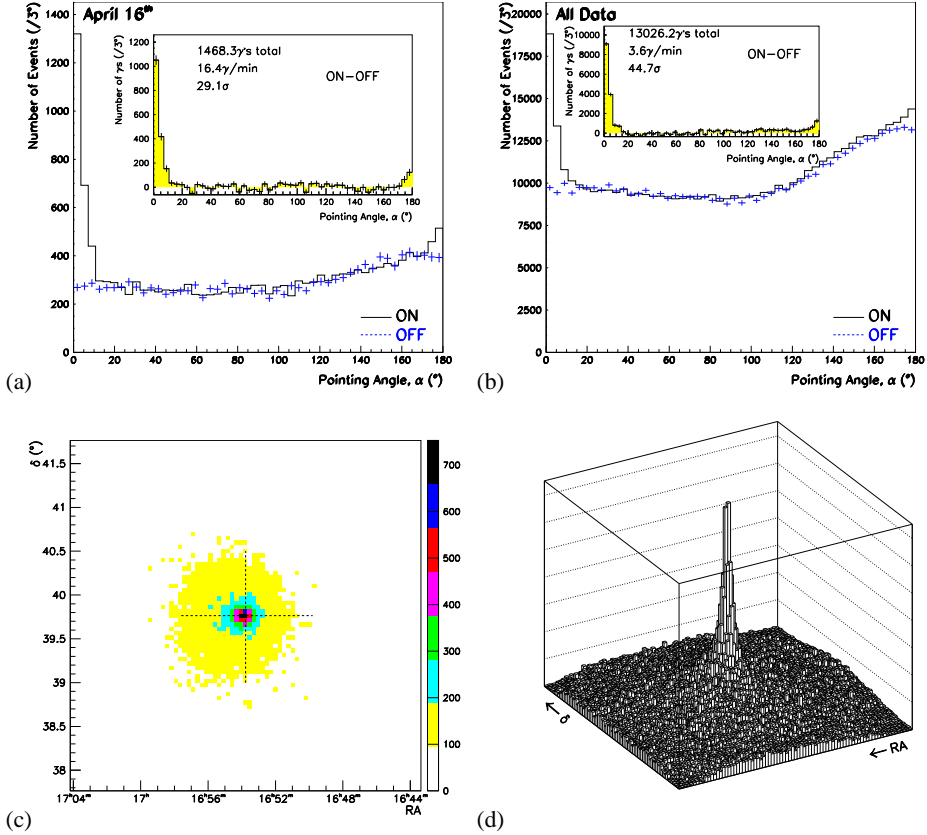

\vbox{
\hbox{
(a)
\epsfig{file=8889.f2a,width=5.5cm,clip=,
bbllx=15,bblly=10,bburx=550,bbury=510}
\hspace{5pt}
(b)
\epsfig{file=8889.f2b,width=5.5cm,clip=,
bbllx=15,bblly=10,bburx=550,bbury=510}
}
\vspace{10pt}
\hbox{
(c)
\epsfig{file=8889.f2c,width=5.5cm,clip=,
bbllx=15,bblly=10,bburx=550,bbury=510}
\hspace{5pt}
(d)
\epsfig{file=8889.f2d,width=5.5cm,clip=,
bbllx=15,bblly=10,bburx=525,bbury=510}
}
\vspace{-6.5cm}
} 
\hfill\parbox[b]{5.3cm}{
\caption{ The Mkn~501 signal seen with the C{\small AT} imaging telescope: 
a), b) $\alpha$ plots for April $16^\mathrm{th}$ and for all data 
respectively ({\small ON} data shown by solid lines, 
{\small OFF} by points with error bars, {\small ON-OFF} inset).
The signal is clearly seen in the direction of the source (small $\alpha$), 
though the direction of some $\gamma$-rays is mis-identified, giving a small 
signal at $\alpha\sim180\degr$.
c), d) two representations of the distributions of the reconstructed angular 
origins for all data 
(the scales in both plots are identical, the bin size 
is $0.05\degr$, no background subtraction has been performed). 
The cross in c) marks the position of Mkn~501. 
Events are selected by the $P_{\chi^2}$ (i.e., shape) cut only.
}
\label{figsigplots}
}
\end{figure*}

A total of 95.3 hours of data were taken on Mkn~501 ({\small ON}),
together with  
36 hours on control  regions ({\small OFF}), between March and October 1997. 
The data used for the analysis were selected by quantitative criteria including
clear weather and stable detector operation.
The data-set has been further limited to zenith angles
$< 44\degr$, 
for which the detector calibration studies have been completed. 
This leaves $57.2^\mathrm{h}$ {\small ON} and $22.5^\mathrm{h}$ {\small OFF} 
data for the analysis presented in the following sections.
Fig.~\ref{figlctot} shows the nightly flux levels as sampled by C{\small AT}
above $250\:\mathrm{GeV}$ between March and October, 1997. The average
flux was approximately twice that of the Crab Nebula, while the
highest flare, recorded on April 
$16^\mathrm{th}$ ($\gamma$-ray rate $\sim\!\!\!16\:\gamma\mathrm{/min}$), reached
about 8 times the Crab level with signal-to-background ratios of 
2.7 for the flare and 0.68 on average  
(see Fig.~\ref{figsigplots}{\small (}a,b{\small )}).

\section{Spectra}
\subsection{Analysis Method}

As stated in Sect.~3, observations consist of {\small ON} and {\small
OFF} source runs, the latter being used to estimate the background due to
hadronic showers. Since the threshold energy of the telescope increases with 
zenith angle, {\small ON} and {\small OFF} data are classified within zenith 
angle bins (with a width of 0.05 in $\cos\theta_z$). Energy bins have been
defined so as to be larger than the telescope's energy resolution.
Trigger and cut efficiencies -- yielding the effective detection area
$\mathcal{A}_\mathrm{eff}(E)$ -- as well
as the energy-resolution function are 
determined by detailed Monte-Carlo simulations of the telescope response
as a function of energy and zenith angle. These simulations have been
checked and calibrated on the basis of several observables using
the nearly-pure $\gamma$-ray signal from the highest flare and muon rings.
This will be reported in a forthcoming paper.

The spectra are derived in two steps. First, the number of events passing the
cuts within
each energy and zenith-angle band is determined for {\small ON}
and {\small OFF} runs. Then, for a given hypothesis on the spectral shape, a
maximum-likelihood estimation of the spectral parameters is performed, taking
into account the effective area and the energy-resolution function of the
telescope. 

Two hypotheses have been assumed for the spectral shape: a simple
power law ($\mathcal{H}_0$) and a curved shape ($\mathcal{H}_1$). In
fact, as we will see below, the data are not compatible with the former
shape. The curved shape is suggested by general
considerations on emission processes within blazar jets (see Sect.~7.1).
Whipple Observatory has also reported a curved shape in terms of a
quadratic law in a  $\log\phi$ {\it vs.} $\log E$ representation
(\cite{samuelson98}). We will use the same framework in order to allow  
for simple comparisons.

To give an estimate of the
relevance of $\mathcal{H}_0$ with respect to $\mathcal{H}_1$, the
likelihood ratio of the 
two hypotheses is defined as $ \lambda = -2 \times \log
\frac{\mathcal{L}(\mathcal{H}_0)}{\mathcal{L}(\mathcal{H}_1)} $. It
behaves (asymptotically) like a $\chi^2$ distribution with one 
degree of freedom.

\subsection{Results}

\begin{table*}[!ht]
\begin{center}
\caption{Data-subset definition as a function of source intensity. 
The differential flux is given as
$\mathrm{d}\Phi/\mathrm{d}E_\mathrm{TeV}= 
\phi_0 E_\mathrm{TeV}^{-(\alpha + \beta\log_{10}\!E_\mathrm{TeV})}$ 
in units of $10^{-11}\:\mathrm{cm^{-2}s^{-1}TeV^{-1}}$.
The fit has been done from $330\:\mathrm{GeV}$ to $13\:\mathrm{TeV}$.
$\phi_0^\mathrm{pl}$ and $\alpha^\mathrm{pl}$ correspond to a simple 
power-law hypothesis ($\mathcal{H}_0$). $\lambda$ is the likelihood ratio 
of the two hypotheses $\mathcal{H}_0$ and $\mathcal{H}_1$.
The last two columns give the peak-emission energies 
$E_\mathrm{max}=10^\frac{2-\alpha}{2\beta}$ (in TeV)
for the second hypothesis, where for the final column, 
the values were obtained by re-fitting with $\beta$ fixed at 0.5.
}
\label{tabspresults}
\begin{tabular}{llcccccccc}
\hline
\noalign{\smallskip}
Set & ON  & $\phi_0^\mathrm{pl}$ &$\alpha^\mathrm{pl}$ & $\phi_0$
&$\alpha$ & $\beta $& $\lambda $& $E_\mathrm{max}$& 
                $E_\mathrm{max}^{\beta=0.5}$\\ 
 & (h)   &  &  & &  &  &  & (GeV) & (GeV) \\ 
\noalign{\smallskip}
\hline
\hline
\noalign{\smallskip}
LF&  13.6 & $\;\:2.72 \pm 0.13$ & $2.45 \pm 0.05 $ 
  & $\;\:3.13 \pm 0.19$ & $2.32 \pm 0.09$& $0.41 \pm 0.17$&
10.7 & $410 \pm 201 $ & $521 \pm 82$\\
MF&  40.5 & $\;\:4.10 \pm 0.10$ & $2.46 \pm 0.03$  
  & $\;\:4.72 \pm 0.14$ & $2.25 \pm 0.05$ & $0.52 \pm 0.08$ & 
   47.1& $583\pm 104$& $563\pm 42$ \\
HF& $\;\:3.1$ & $14.5\;\: \pm 0.50$ & $2.21 \pm 0.03$ 
  & $17.6\;\: \pm 0.61$ & $2.07 \pm 0.04$ & $0.45 \pm 0.09$ &
  29.1 & $840\pm 108$ & $890\pm77$ \\
\hline
AV&  57.2 & $\;\:4.56 \pm 0.10$ & $ 2.46\pm 0.02$ 
  & $\;\:5.19 \pm 0.13$ & 
  $2.24 \pm 0.04$ & $0.50 \pm 0.07$ & 61.5&  $578\pm\;\; 98$& $589\pm 38$ \\
\noalign{\smallskip}
\hline
\end{tabular}
\end{center}
\end{table*}

In order to investigate spectral variations as a function of source
intensity, three independent data-subsets corresponding to different activity
states of Mkn~501 -- high-intensity flares
(HF, integral flux $> 53 \times 10^{-11}\:\mathrm{cm^{-2}s^{-1}}$
above $250\:\mathrm{GeV}$), 
low-intensity runs
(LF, integral flux $<12 \times 10^{-11}\:\mathrm{cm^{-2}s^{-1}}$),
and mid-intensity flares (MF) -- have been established. The 
average over the complete data-set (AV) has also been considered.
The quadratic law used to estimate the spectral parameters is
$\phi_0 E_\mathrm{TeV}^{-(\alpha + \beta\log_{10}\!E_\mathrm{TeV})}$.
Results for each data-subset are given in Table~\ref{tabspresults}.
Only statistical errors are quoted in this table.

As noted in Sect.~2, the systematic errors on the absolute energy
scale, conservatively estimated at 20\%, lead to an
error on the absolute flux depending on the spectral index 
(referred to below as ``sys-atm''). 
Additional errors coming from the limited statistics of the
simulated $\gamma$-rays 
(referred to below as ``sys-MC'') 
used to determine $\mathcal{A}_\mathrm{eff}(E)$
are twofold:
\begin{list}{--}{\topsep 0pt}
\item an overall scaling error which affects the absolute flux.
\item possible
distortions of $\mathcal{A}_\mathrm{eff}(E)$ as a function of energy
within the Monte Carlo errors affecting the parameter
$\alpha$. The curvature parameter $\beta$ is, however, almost
unaffected. 
\end{list}
Both parameters have proved to be insensitive to modifications of 
the analysis procedure (which modify efficiencies and energy
resolution functions), showing that they are robust.

For the AV spectrum, the complete results are:

\indent $\phi_0 = \mathrm{ 5.19 \pm 0.13^{stat} \pm 0.12^{sys-MC} }\\
\indent \indent \indent \;\;\;\;
\stackrel{+1.66}{_{-1.04}}^\mathrm{sys-atm}
\times 10^{-11}\:\mathrm{cm^{-2}s^{-1}TeV^{-1}}$, \\
\indent$\alpha \:\, = 2.24 \pm \mathrm{ 0.04^{stat} \pm 0.05^{sys} }$, and\\
\indent$\beta \hspace{4.2pt} = 0.50 \pm 0.07^\mathrm{stat}$ (negligible systematics). \\
\indent The corresponding covariance matrix for the fit, based on the 
statistical errors, is: \\
\indent $\left( \begin{array}{rrr}
 0.01767 & -0.00052 &  0.00443 \\
-0.00052 &  0.00168 & -0.00203 \\
 0.00443 & -0.00203 &  0.00475 \end{array} \right),$\\
with $\alpha$ and $\beta$ being correlated.

The likelihood ratio, $\lambda$, for $\mathcal{H}_0$ and $\mathcal{H}_1$ 
shows that the HF and MF states (and the AV spectrum) 
have significantly curved shapes:  $\lambda$ has values of
29.1 and 47.1 respectively, corresponding to chance probabilities of
$7\times 10^{-8}$ and $7\times 10^{-12}$ of accepting $\mathcal{H}_1$ were
the power-law hypothesis $\mathcal{H}_0$ true. 
For the LF state, the curvature significance is somewhat lower
(a probability of $10^{-3}$), but this 
could be attributed to the weak
intensity of the source.

\begin{figure}[tbh]
\leavevmode
\begin{center}
\hbox{\hspace{5pt}
\epsfig{file=8889.f3,width=8.5cm,clip=,
bbllx=10,bblly=20,bburx=546,bbury=480}
}
\caption{
$\nu F_\nu$ (or $E^2\,\mathrm{d}\Phi/\mathrm{d}E$)
spectra for the HF, MF, and LF states. 
Curves represent the global flux estimated by method described in Sect. 4.1.  
The points, corresponding to the individual bin intensities, are only
indicative and were obtained by the following procedure: 
The expected number, $N_\mathrm{fit}$, of $\gamma$-ray events in each energy bin 
is computed using the fitted flux parameters and taking into account the 
acceptance and energy resolution functions of the telescope.
The observed number of $\gamma$-ray events, $N_\mathrm{obs}$, is obtained by 
subtracting {\small ON} and {\small OFF} data.
The ratios $\frac{N_\mathrm{obs}}{N_\mathrm{fit}}$ give a good 
approximation of the fluctuation in each bin, and are used to correct
the flux values yielded by the fitted curve.
Error bars give the combination of uncertainties on $N_\mathrm{obs}$ and on
the effective detection area (statistical only).}
\label{fignfnall}
\end{center}
\end{figure}

The SEDs are shown in
Fig.~\ref{fignfnall}: as can be seen, the VHE peak emission of Mkn~501 
clearly takes place in the range of several hundred GeV. 
The corresponding maximum energy, $E_\mathrm{max}$,
of the equivalent quadratic form
-- proportional to 
$E_\mathrm{TeV}^{-2}(E_\mathrm{TeV}/E_\mathrm{max})^
{-\beta\log_{10}(E_\mathrm{TeV}/E_\mathrm{max})}$ -- is given by
$E_\mathrm{max} =10^\frac{2-\alpha}{2\beta}$
(see Table~\ref{tabspresults}).

\begin{table*}[bt]
\begin{center}
\caption{Hardness-ratio values for five different flux bands 
(zenith angle $< 25\degr$). 
The average integral flux per band above $250\:\mathrm{GeV}$ is given in units of 
$10^{-11}\:\mathrm{cm^{-2}s^{-1}}$ and has been computed 
assuming the average spectral shape. Set I corresponds to the highest
flare observed on April $16^\mathrm{th}$ and set II to  
the second-highest flares of April $13^\mathrm{th}$ and June $7^\mathrm{th}$.}
\label{tabhr}
\begin{tabular}{lrcccc}
\hline
\noalign{\smallskip}
Set & {\small ON} (h) & Flux & 
       $R_{[>900/>450]}$ &$R_{[>1200/>600]}$ & $R_{[>1500/>900]}$\\ 
\noalign{\smallskip}
\hline
\hline
\noalign{\smallskip}
I   &   1.5 & ~$\,$68--105 &$0.56 \pm 0.02$& $0.50 \pm 0.02$ &$0.56 \pm 0.03$\\
II  &   3.5 & ~45--68~     &$0.53 \pm 0.02$& $0.49 \pm 0.02$ &$0.56 \pm 0.02$\\
III &   5.5 & ~28--45~     &$0.47 \pm 0.02$& $0.44 \pm 0.02$ &$0.53 \pm 0.03$\\
IV  &  10.5 & ~17--28~     &$0.48 \pm 0.02$& $0.42 \pm 0.02$ &$0.49 \pm 0.02$\\
V   &  20.5 & 5.5--17~ &$0.45 \pm 0.02$& $0.41 \pm 0.02$ &$0.51 \pm 0.03$\\
\noalign{\smallskip}
\hline
\end{tabular}
\end{center}
\end{table*}

\begin{figure*}[bt]
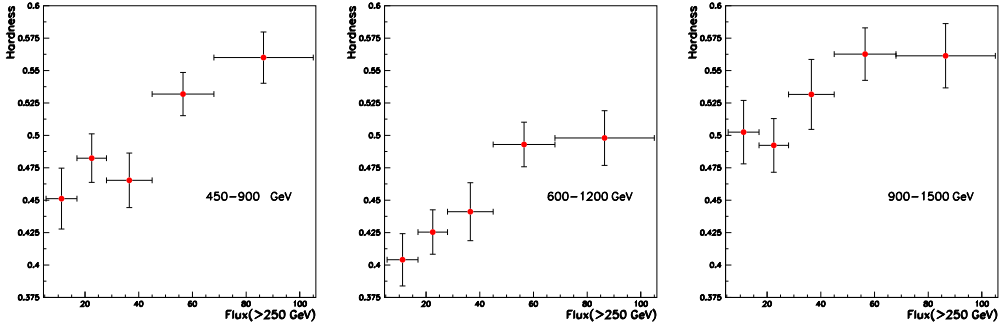

\vbox{
\epsfig{file=8889.f4a,width=4.2cm,
bbllx=60,bblly=50,bburx=520,bbury=515}
\hspace{5pt}
\epsfig{file=8889.f4b,width=4.2cm,
bbllx=60,bblly=50,bburx=520,bbury=515}
\hspace{5pt}
\epsfig{file=8889.f4c,width=4.2cm,
bbllx=60,bblly=50,bburx=520,bbury=515}
\vspace{-2.5cm}
} 
\hfill\parbox[b]{4.2cm}{
\caption{Hardness-ratio values {\it vs.} source intensity for three
  different energy bands. Intensities are given as the
  integral flux above $250\:\mathrm{GeV}$ in units of 
  $10^{-11}\:\mathrm{cm^{-2}s^{-1}}$.
}
\label{fighr} 
}
\end{figure*}

There is some indication
of an increase in $E_\mathrm{max}$ with source intensity, which is 
equivalently shown by the variation of the fitted values of $\alpha$.
Although the power-law hypothesis is clearly not favoured here, the
corresponding variation of $\alpha^\mathrm{pl}$ values follows the
same tendency. 
On the other hand, the curvature (given by $\beta$) is
compatible with being constant for the three data-subsets. 
Fixing $\beta=0.50$ yields
$E_\mathrm{max}$ with smaller errors, as shown in the last column
of Table~\ref{tabspresults}.
This reinforces a possible correlation
between intensity and $E_\mathrm{max}$.
We will investigate this question in more detail in the next section
by looking at intensity-hardness correlations.

In order to determine the highest-energy $\gamma$-rays measured, the effects
of the energy-resolution function and the uncertainty in the absolute energy 
scale must be taken into account.  There is a signal at the
$2\sigma$ significance level above $12\:\mathrm{TeV}$, after correction for the 
number of lower-energy events which migrate above this energy due to the 
energy resolution (and also to the lesser number which migrate to lower energy).
Taking into account also the uncertainty in the absolute energy 
scale implies (conservatively) that there is a $2 \sigma$ signal 
above $10\:\mathrm{TeV}$.

\section{Spectral Hardness {\it\bf vs.} Source Intensity}

As compared to a usual spectrum analysis method, a more
robust manner to check for spectral changes is to consider two given
energies $E_\mathrm{low}$ and $E_\mathrm{mid}$, and 
the hardness ratio defined as: $R = \frac{N_{E>E_\mathrm{mid}}}
{N_{E>E_\mathrm{low}}}$. 

Given the large energy-bins, this ratio is rather insensitive to
statistical fluctuations and entails the estimation of only one  
parameter. 
Nevertheless, as the detection threshold of atmospheric \v Cerenkov
telescopes increases with zenith angle and is sensitive to slight
changes in 
weather conditions, it is desirable to limit the data-set to low
zenith angle runs (here $25\degr$) and 
choose $E_\mathrm{low}$ well above the telescope trigger threshold of
$250\:\mathrm{GeV}$ (here $450\:\mathrm{GeV}$ or above). 

Although these two restrictions reduce the available statistics, 
they allow for minor acceptance corrections by weighting data to
normalize them to those taken at the Zenith. This minimizes any
possible systematic effects. Furthermore, to achieve the best energy
resolution, an additional cut on shower impact parameter 
($< 130\:\mathrm{m}$ at the Zenith) has been introduced.

The limited statistics still forbid a reliable computation of the
hardness ratio on a run-by-run (or 
even night-by-night) basis, except for the highest-intensity runs.
Therefore, data have been divided into five sets with different average
fluxes. For each data-set in flux, the hardness ratio has been computed
for three different energy bands,
$[E_\mathrm{low},E_\mathrm{mid}]_\mathrm{GeV}
=\{[450,900];[600,1200];[900,1500]\}$,
hereafter referred to as $R_{[>900/>450]}$, $R_{[>1200/>600]}$, and
$R_{[>1500/>900]}$. The number of hours for each data-set and the corresponding average
flux and hardness ratio values are given in Table~\ref{tabhr}.

As can be seen in Fig.~\ref{fighr}, a clear correlation between
intensity and hardness is present for all energy-band definitions.
The chance probabilities of observing such an effect in the case of 
an absence of correlation for $R_{[>900/>450]}$, 
$R_{[>1200/>600]}$ and $R_{[>1500/>900]}$ are respectively: 
$3.8\times 10^{-4}$, $9\times 10^{-4}$, and $7\times 10^{-2}$. We note that
as $E_\mathrm{low}$ and $E_\mathrm{mid}$ are raised to higher
energies, the significance of the correlation decreases owing to lack of
statistics.

\begin{table*}[tbh]
\begin{center}
\caption{Observation logs and power output 
(in units of $10^{-11}\:\mathrm{erg\cdot cm^{-2}s^{-1}}$) 
for BeppoS{\small AX} 
(\cite{pian98}) and C{\small AT}. 
For C{\small AT}, the integral fluxes corresponding to April $7^\mathrm{th}$ 
and $11^\mathrm{th}$ are calculated assuming the spectral shape from the LF 
and MF data-subsets, respectively.
}
\label{tabobscatsax}
\begin{tabular}{l|cc|cc}
\hline
\noalign{\smallskip}
Date & BeppoS{\small AX}& X-ray power output 
& C{\small AT} & VHE power output  \\ 
& Observing time (UT) & 13--$200\:\mathrm{keV}$ 
& Observing time (UT) & 0.25--$15\:\mathrm{TeV}$ \\
\noalign{\smallskip}
\hline
\hline
\noalign{\smallskip}
April $7^\mathrm{th}$  & 
  05:30:29 $\rightarrow$ 16:01:47& 
    $\;\:37.5 \pm 1.5 $& 
    00:34:06 $\rightarrow$ 03:56:53 &
    $\;\;\;\,  9.8 \pm  2.4$ \\
April $11^\mathrm{th}$ & 
 05:53:57 $\rightarrow$ 16:25:49& 
    $\;\:51.5 \pm 1.5 $& 
    01:06:57 $\rightarrow$ 03:53:16 &
    $\;\: 22.4 \pm  3.1$ \\
April $16^\mathrm{th}$ & 
 03:36:52 $\rightarrow$ 14:35:48& 
    $158.0 \pm 2.0$& 
    01:42:34 $\rightarrow$ 03:45:55 &
    $100.0 \pm 7.5$ \\
\noalign{\smallskip}
\hline
\end{tabular}
\end{center}
\end{table*}

Considering the curved spectral shape measured in the previous section,
and fixing the curvature at $\beta = 0.50$, we can calculate the shift in 
$E_\mathrm{max}$ necessary to cause the observed maximum change in $R$. 
The variation is of the order of $400\:\mathrm{GeV}$,
which is fully compatible with the results of the 
previous section seen in the final column of Table~\ref{tabspresults}. 

\section{Hard X-ray and $\gamma$-ray correlation}

During the April 1997 multi-wavelength campaign, Mkn~501 was observed from 
X-ray to VHE energies (e.g., see \cite{catanese97}). 
In the energy range 0.1--$200\:\mathrm{keV}$, the BeppoS{\small AX} satellite
observed for three periods of $\sim\!\!\!10$ hours: on April $7^\mathrm{th}$,
$11^\mathrm{th}$ and $16^\mathrm{th}$ (\cite{pian98}). 
The RXTE satellite observed 
in the 2--$80\:\mathrm{keV}$ range (\cite{lamer98}) between July $11^\mathrm{th}$
and $16^\mathrm{th}$. X-ray and high energy $\gamma$-ray observations were
also performed by the O{\small SSE} and E{\small GRET} instruments
respectively, on board
CGRO. In the VHE regime, C{\small AT}, H{\small EGRA}, Whipple and other
instruments were active (\cite{protheroe97}).

In the X-ray range, extremely hard spectra were reported by BeppoS{\small AX}
indicating that the synchrotron power output peaked at $100\:\mathrm{keV}$ or 
more. O{\small SSE} observations gave similar conclusions (\cite{catanese97}). 

\begin{figure}[b]
\leavevmode
\begin{center}
\epsfig{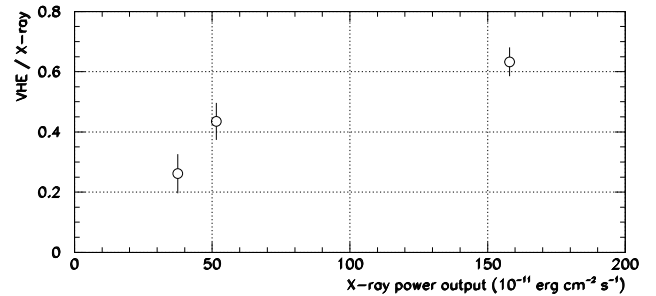}
\caption{
The ratio of VHE to X-ray power outputs as a function of the X-ray output,
from the data in Table~\ref{tabobscatsax}. On this plot, a linear relation 
between the VHE and X-ray intensity would appear as a horizontal line,
and a quadratic relation as a line passing through the origin.
}
\label{figrapcatsax} 
\end{center}
\end{figure}

The highest-intensity TeV flare recorded by C{\small AT} in 1997 
occurred on April $16^\mathrm{th}$ (see Fig.~\ref{figlctot}),
where C{\small AT} and BeppoS{\small AX} observations overlapped for about
ten minutes, providing some simultaneous VHE $\gamma$-ray and X-ray data. 
The observation logs are given in Table~\ref{tabobscatsax}. 
On April $7^\mathrm{th}$ and $11^\mathrm{th}$, the source was less bright and
only quasi-simultaneous data were available, as there was a delay of about two
hours between C{\small AT} and BeppoS{\small AX} pointing periods. 
However, data from all \v Cerenkov telescopes (C{\small AT},
Whipple and H{\small EGRA}) show no large variations in source flux for the 
periods $5^\mathrm{th}$ to $9^\mathrm{th}$ and $10^\mathrm{th}$ to
$12^\mathrm{th}$ (\cite{protheroe97}). Hence, $\gamma$-ray and X-ray
levels can be considered as simultaneous data and compared in a
reliable manner.

Table~\ref{tabobscatsax} shows comparisons of the VHE and X-ray power
output from BeppoS{\small AX} and C{\small AT} data. 
The ratio of VHE to X-ray power outputs is plotted
in Fig.~\ref{figrapcatsax} as a function of the X-ray power, and
clearly increases with source brightness. 

\begin{figure}[bt]
\leavevmode
\begin{center}
\epsfig{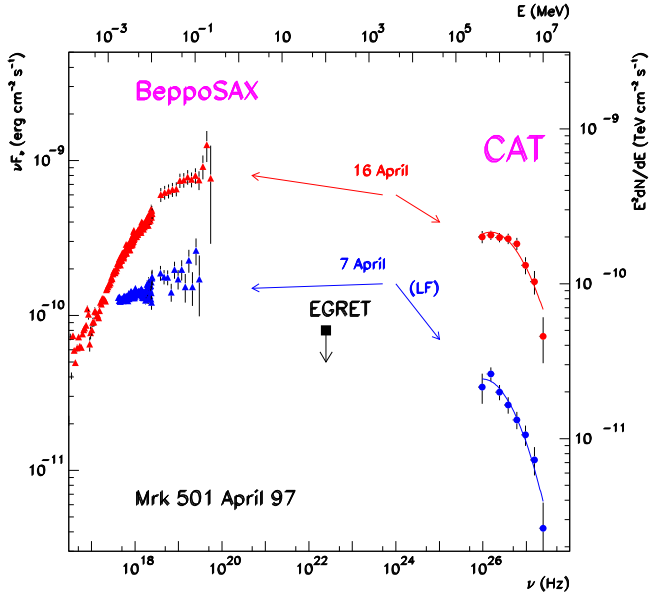}
\caption{X-ray and VHE spectra given as
$\nu F_\nu$. For April $7^\mathrm{th}$ and $16^\mathrm{th}$, BeppoS{\small AX}
data are from Pian et al. (1998). 
The LF subset in C{\small AT} was used to represent the spectrum corresponding
to  April $7^\mathrm{th}$.
The EGRET upper limit is taken from
Samuelson et al. (1998) and corresponds to observations between
April $9^\mathrm{th}$ and $15^\mathrm{th}$.
}
\label{fignfncatsax} 
\end{center}
\end{figure}

The spectra are shown in Fig.~\ref{fignfncatsax}.
Because of the very high intensity of the source on April $16^\mathrm{th}$, 
it is possible to derive its spectrum for that single observation (one
and a half hours of data).  
In order to represent the spectrum corresponding to April $7^\mathrm{th}$, 
the LF data-subset comprising runs with similar flux levels (within 30\%) 
was used.

Four characteristics come out of these comparisons: 
\begin{list}{--}{\topsep 3pt}
\item[(i)]
the fluxes and SED shapes seen in the C{\small AT} and BeppoS{\small AX} data, 
together with the E{\small GRET} upper limit in the GeV domain, 
imply a double structure in the power output of Mkn~501;
\item[(ii)]
hard X-rays and VHE $\gamma$-rays show correlated variability;
\item[(iii)] 
the luminosity in X-rays remains higher than in $\gamma$-rays;
\item[(iv)]
during the flares, the VHE intensity recorded by C{\small AT}
increased by a larger factor than the X-ray intensity recorded by 
BeppoS{\small AX} (Table~\ref{tabobscatsax}).
Therefore, the ratio of VHE to X-ray intensities increased with
source brightness (Fig.~\ref{figrapcatsax}).
\end{list}

\section{Discussion}

C{\small AT} observations of Mkn~501 during 1997 have revealed several
spectral properties: 
a curved shape and an emission extending above $10\:\mathrm{TeV}$ 
(Sect.~4), hardness--intensity correlations (Sect.~5), X-ray and $\gamma$-ray
correlations (Sect.~6).
Together with the power deficit in the GeV range,
this strongly suggests a two-component emission spectrum for Mkn~501. 

It is generally accepted that the high luminosity of
blazars ($\sim\!\!\!10^{45-48}$ erg/s) requires a relativistically-amplified
beamed-emission. Beaming is 
also required for the source to be transparent to $\gamma$-rays if
both VHE and IR to UV emissions originate in the same region 
of the jet (\cite{gehrels95}; \cite{dondi95}).  
Two major processes have been suggested as the dominant source
of the high-energy emission
from blazars: inverse Compton (IC) scattering of low-energy photons by 
ultra-relativistic electrons, 
and $\pi^{0}$-decay through photo-production or secondary 
production of pions by an ultra-high-energy hadronic component in the jet
(\cite{mannheim92}; \cite{mannheim93}).

In the IC case, a double-structure spectrum in terms of synchrotron
and IC emissions is predicted with the two components being correlated 
in intensity, hardness, energy of the peak luminosities, etc. The
correlations depend on the nature of the 
target-photon population. This may consist of either
synchrotron photons within the jet -- the synchrotron self-Compton,
or SSC, mechanism (\cite{konigl81}; \cite{marscher85}; \cite{ghisellini96}) --
or external radiation, either from the accretion disk 
(\cite{sikora94}; \cite{dermer92}), or
reflected by the broad-line region clouds
(\cite{ghisellini-madau96}).
Spectral variations of the primary electrons (due for example to a change in
the injection-acceleration efficiency or in cooling conditions), seen
through their effect on the synchrotron radiation, should be reflected in the
high-energy emission, especially when the IC interactions take place in
the Thomson limit. In the Klein-Nishina regime, hardening effects can be
attenuated because of a severe cross-section drop at high energies.

In hadronic models, no particular predictions are made for a
marked two-component emission. A continuum emission is
predicted due to secondary cascades produced by
hadronic interactions of ultra-high-energy protons, which tend to fill
the gap between the high-energy 
($\gamma$-ray) and low-energy (X-ray) domains.

\subsection{Curved Shape}

The spectral shape derived in Sect.~4 between $330\:\mathrm{GeV}$
and $13\:\mathrm{TeV}$ is clearly curved.
This curvature is in contrast with reports on the other TeV blazar, Mkn~421,
whose spectrum has been observed to be compatible with a power-law
(\cite{whipplemrk421}; \cite{catmrk421}). 
The curved shape is predicted by IC models of blazar spectra but can
only be seen if the IC
component is observed over a broad dynamic range close to the
maximum power emission energy (i.e., close to the maximum of the IC bump 
in a $\nu F_\nu$ representation). The curvature could arise from
intrinsic absorption effects (within the jet or due to nearby radiation fields,
from the accretion disk, for example) and/or from a cutoff in the primary 
particle spectrum. However, when observing within a limited energy range -- 
as compared to the energy scale over which the emission phenomenon is taking 
place -- not including the peak emission region, a power-law spectrum is 
naturally seen.  We note that the reported synchrotron peak of Mkn~501 during 
1997 is well above that of Mkn~421 ($\sim\!\!\!100\:\mathrm{keV}$ as compared 
to $\sim\!\!\!1\:\mathrm{keV}$) and that the IC and synchrotron peak-emission 
energies are expected to be correlated in IC schemes. This correlation has 
also been established on a purely observational basis using a large sample of 
blazars (\cite{fossati98}).
Thus, the difference in observed spectral shapes for Mkn~421 and Mkn~501
may well arise because the peak of the high-energy component for Mkn~421
is below the VHE energy domain, whereas the peak for Mkn~501 is well
above our threshold. 

The overall spectrum, and particularly the highest energy $\gamma$-rays
(10--$20\:\mathrm{TeV}$), may be affected by intergalactic absorption due to 
pair creation on the 
intergalactic background IR radiation (\cite{stecker98}; \cite{biller98}). 
Therefore any interpretation of VHE spectra in terms of emission
models should take 
this effect into account. It could be disentangled from intrinsic
absorption features if we dispose of a large sample of VHE
emitters. Nevertheless, the
different spectral shapes of Mkn~501 and Mkn~421 -- which have similar
red-shifts
and have been detected above $10\:\mathrm{TeV}$ (\cite{whipplemrk421})
-- imply already that the curvature of the former must be
mainly due to intrinsic features.

\subsection{Hardness--intensity and $\gamma$-ray/X-ray Correlations}

In Sect.~5, a correlation between VHE intensity and spectral hardness
was demonstrated. If we assume a constant curvature for 
different-activity states as indicated by the data, this
effect is equivalent to an increase of approximately $400\:\mathrm{GeV}$ for 
the peak emission energy $E_{\mathrm{max}}$ for an intensity change of about
an order of magnitude between the LF and HF subsets. 

BeppoS{\small AX} results have established that the energy of the maximum
power output during the active period is much larger than during the 
low-emission state (\cite{pian98}). A comparison of April $7^\mathrm{th}$ 
and $16^\mathrm{th}$ X-ray data suggests that this hardening 
continues as the intensity increases during the flare.

This simultaneous observation of flaring activity and spectrum hardening,
in both X-ray and VHE energy ranges, is naturally explained by a
common primary  
electron population, as expected in the framework of inverse Compton
models.
Pian et al. (1998) claim that a significant change in the magnetic field
intensity or the Doppler factor is unlikely since the expected changes in
the lower-energy part of Mkn~501 spectrum which would result have not  
been observed; i.e., the X-ray emission below $0.5\:\mathrm{keV}$ did not show 
any dramatic variation. In this case, this would leave the injection of a 
transient electron population and/or a change in the acceleration efficiency 
(an increase of the maximum energy of electrons) as the most
likely source of the flares.

In the framework of IC models, the observed increase of the VHE to X-ray 
intensity ratio with source brightness (see Fig.~\ref{figrapcatsax}) is
qualitatively in agreement with the SSC mechanism which predicts a
quadratic increase of VHE intensity with X-ray brightness
due to the fact that both the electron population and the ``target"
synchrotron populations are enhanced.
The yield of the SSC contribution to the IC process should be damped
by the decreasing cross-section in the extreme relativistic Klein-Nishina
regime. The highest intensity data point in Fig.~\ref{figrapcatsax}
is consistent with such an effect.

\subsection{Comparison with other VHE detections}

The Whipple group's results for Mkn~501 between $300\:\mathrm{GeV}$ and
$10\:\mathrm{TeV}$ (\cite{samuelson98}) show an average 
spectrum with $\alpha = \mathrm{ 2.20 \pm 0.04^{stat} \pm 0.05^{syst} }$ and
$\beta = \mathrm{ 0.47 \pm 0.07^{stat} }$, in good agreement with C{\small AT} 
results within the statistical errors. 
H{\small EGRA} (\cite{aharonian99}) has also demonstrated a curved
spectral shape. 
Given the source variability and the different source sampling due to local
conditions, one does not expect the absolute flux levels observed in the 
three experiments to be identical. 
Nonetheless, the convergence of these three observations
as regards the spectral {\sl shape} is noteworthy.

No hardening effect has been claimed by either Whipple or H{\small EGRA}. 
While this appears to 
contradict the effect seen in our data, it may be explained for each
case. We note that Whipple's reported results
are limited to the period between 
February and June 1997 and include $21.1^\mathrm{h}$ of data. 
The C{\small AT} results
are based on a larger sample ($40.5^\mathrm{h}$ of observations for the
hardness-intensity studies), 
and include runs from June to October 1997, providing large statistics 
of the source emission in a lower-activity state. The effect reported
here could hardly be seen with lower statistics.

Considering H{\small EGRA} results, their higher energy threshold
might well result in a lower sensitivity to the hardness-intensity
correlation: as shown in Sect.~6, this effect becomes less
significant when we consider higher energy events.

\section{Conclusion}

The results reported in this paper are based on the large sample of data taken
by C{\small AT} from March to October 1997, including periods with  very
different source intensities.
The Mkn~501 spectrum has been measured from $330\:\mathrm{GeV}$ to 
$13\:\mathrm{TeV}$. The curvature reported by the Whipple group
is confirmed.  
In addition, C{\small AT} data show an intensity-hardness correlation which
can be simply described by a shift of the peak TeV emission energy.

The overall picture emerging from these spectral properties and from
the comparison with simultaneous and 
quasi-simultaneous BeppoS{\small AX} observations favours the  
interpretation in terms of an energetic electron beam propagating in 
a magnetic plasma and producing X-rays through synchrotron radiation
as well as VHE $\gamma$-rays through inverse 
Compton scattering of low-energy photons.

\begin{acknowledgements}
The authors wish to thank the French national institutions IN2P3/CNRS
and DAPNIA/DSM/CEA for supporting and funding the C{\small AT} project.
The C{\small AT} telescope was also partly funded by the
Languedoc-Roussillon region and the Ecole Polytechnique. The authors
also wish to thank Electricit\'e de France for making available to them
equipment at the former solar plant ``Th\'emis'' and allowing the 
building of the new telescope and its hangar. They are grateful to the
French  and Czech ministries of Foreign Affairs for providing grants
for physicists' travel and accommodation expenses.
\end{acknowledgements}

\end{document}